\documentclass[twocolumn,english,aps,pra,showpacs,superscriptaddress,]{revtex4}
\usepackage[T1]{fontenc}
\usepackage[latin9]{inputenc}
\usepackage{amsmath}
\usepackage{amssymb}
\usepackage{graphicx}
\usepackage{esint}
\usepackage{color}
\usepackage[colorlinks,citecolor=blue]{hyperref}

\makeatletter
\@ifundefined{textcolor}{}
{%
 \definecolor{BLACK}{gray}{0}
 \definecolor{WHITE}{gray}{1}
 \definecolor{RED}{rgb}{1,0,0}
 \definecolor{GREEN}{rgb}{0,1,0}
 \definecolor{BLUE}{rgb}{0,0,1}
 \definecolor{CYAN}{cmyk}{1,0,0,0}
 \definecolor{MAGENTA}{cmyk}{0,1,0,0}
 \definecolor{YELLOW}{cmyk}{0,0,1,0}
}

\usepackage[english]{babel}

\makeatother

\usepackage{babel}
\begin{document}
\title{Towards quantum simulation of Sachdev-Ye-Kitaev model}
\author{Ye Cao}
\affiliation{School of Physics, Beijing Institute of Technology, Beijing 100081, China}
\author{Yi-Neng Zhou}
\affiliation{Department of Physics, Renmin University of China, Beijing 100872, China}
\author{Ting-Ting Shi}
\affiliation{Department of Physics, Renmin University of China, Beijing 100872, China}
\author{Wei Zhang}
\email{wzhangl@ruc.edu.cn}
\affiliation{Department of Physics, Renmin University of China, Beijing 100872, China}
\affiliation{Beijing Key Laboratory of Opto-electronic Functional Materials
and Micro-nano Devices, Renmin University of China, Beijing 100872, China}
\date{\today}
\begin{abstract}
We study a simplified version of the Sachdev-Ye-Kitaev (SYK) model with real interactions 
by exact diagonalization. Instead of satisfying a continuous Gaussian distribution, 
the interaction strengths are assumed to be chosen from discrete values with a finite separation.
A quantum phase transition from a chaotic state to an integrable state is observed by increasing 
the discrete separation. Below the critical value, the discrete model can well reproduce various 
physical quantities of the original SYK model, including the volume law of the ground-state entanglement, 
level distribution, thermodynamic entropy, and out-of-time-order correlation (OTOC) functions. 
For systems of size up to $N=20$, we find that the transition point increases with system size, 
indicating that a relatively weak randomness of interaction can stabilize the chaotic phase. 
Our findings significantly relax the stringent conditions for the realization of SYK model, 
and can reduce the complexity of various experimental proposals down to realistic ranges. 
\end{abstract}
\maketitle

\section{Introduction}
\label{sec:intro}

Since 1970s, models of infinite-range random interacting particles
have been introduced and gained widespread attention in condensed matter
physics~\cite{Sherrington1975,Sachdev1993,Parcollet1999,Georges2000,
Georges2001,Arrachea2002,Camjayi2003}. 
The ground states of these models are believed to break long-range order and
be non-Fermi liquid or spin-glass. Recently, Kitaev has proposed an exactly solvable model~\cite{Kitaev2015} 
equivalent to Sachdev and Ye's proposal~\cite{Sachdev1993} in the sense of 
identical saddle-point equations in the large-$N$ limit. 
This so-called Sachdev-Ye-Kitaev (SYK) model, as defined in Eq.~(\ref{eq:original_syk_hamiltonian}),
springs up in a wide range of areas since it is believed to be a holographic dual of 
the $\text{AdS}_{2}$ horizons of charged black holes and 
intimately related to various important topics, including quantum chaos, 
eigenstate thermal hypothesis (ETH), and information theory, to name a 
few~\cite{Sachdev2015,Maldacena2016,Hosur2016,Garcia2016,
Polchinski2016,You2017,Scaffidi2019,Garcia2018,Garcia2019-chaos}.
Inspired by the prototype, some variant models were proposed to explore (topological) phase transitions~\cite{Altland2019,Guo2019,Zhang2018, Banerjee2017,Bi2017,Can2019-phase-transition}, 
the stability of non-Fermi liquid phase~\cite{Garcia2017,Jian C-.M2017,Jian S.-K2017,Chen2017,Song2017,Trajewski2019},
the properties in high dimensions~\cite{Jian C-.M2017, Jian S.-K2017,Banerjee2017}, as well as level statistics and
many-body localization~\cite{Iyoda2018,Garcia2019-mbl}. 

Alongside the exciting theoretical progresses, the experimental realization of the SYK
model is still hindered mainly by the complicated form of interactions. 
Recently, a few proposals have been put forward in different physical
platforms. In condensed matter systems, there are suggestions using Majorana
zero modes residing on the surface of a topological insulator~\cite{Pikulin017}
or hosted by topological superconducting wires coupled with a quantum
dot~\cite{Chew2017}. Besides, a device without the requirement of
superconductivity is proposed in an irregular shaped graphene flake~\cite{Chen2018}, 
where the transport characteristics are further examined theoretically~\cite{Can2019-transport}. 
Another possible route towards this goal is to employ the high controllability of 
ultracold quantum gases. Danshita {\it et al.} have presented a variant of SYK 
model~\cite{Danshita2017}, which can be realized in principle by coupling two 
confined fermionic atoms with molecular states via photo-association lasers. 
In all these proposals, much effort has been devoted to fulfill the three
conditions of the SYK model: (i) the inter-site tunneling terms are absent, 
(ii) all particles are coupled via an infinitely long-range four-fermion
interaction $J_{ijkl}$, and (iii) the interaction strengths of $J_{ijkl}$ are distributed 
in a Gaussian form. While the tunneling between sites can be turned off rather 
easily in some proposals, the latter two points require very delicate and complex
setups which are too difficult to be implemented in experiments for systems of only a few particles. 
An important question then naturally arises: {\it Whether and to what extent the stringent 
requirement of a perfectly random all-to-all interaction can be relaxed to simulate the SYK model?}

In this work, we consider a significantly simplified variant of SYK model, where the random 
interaction strengths can only take real and discrete values separated by a give distance, 
i.e., $0$, $\pm d$, $\pm 2d$, $\dots$. The probability for the interaction to take a specific value 
renders a Gaussian distribution, hence for large enough separation $d$, the interaction is very 
likely to be zero and the majority of particles become mutually non-interacting. This model 
significantly relaxes the stringent conditions (ii) and (iii) aforementioned, and paves the route
towards the experimental simulation of SYK model by reducing the complexity in a large extent.
Using exact diagonalization, we investigate the level distribution of the spectrum and find 
a quantum phase transition from an SYK-like phase to a Fermi liquid phase by increasing 
the separation $d$ of interaction strength. This transition leaves clear signatures in various
physical properties, including the entanglement entropy of the ground state, 
the thermodynamic entropy at finite temperature, and the out-of-time-ordered correlation (OTOC) function.
For system size up to $N=20$, we find that the transition point increases with $N$, 
which means for larger systems the conditions (ii) and (iii) can be further relaxed 
and be more easily implemented.

\section{Model Hamiltonian}
\label{sec:model}

The SYK model in $0+1$ space-time dimensions is governed by a model
of spinless fermions on a one-dimensional lattice:
\begin{eqnarray}
\mathcal{H}_{\mathrm{SYK}} & = & \frac{1}{(2N)^{\frac{3}{2}}}\sum_{i,j,k,l}^{N}J_{ij,kl}c_{i}^{\dagger}c_{j}^{\dagger}c_{k}c_{l}-\mu\sum_{i}c_{i}^{\dagger}c_{i},\label{eq:original_syk_hamiltonian}
\end{eqnarray}
 where $c_{i}^{\dagger}$ are fermion creation operators and $J_{ij,kl}$
are complex random couplings satisfying a Gaussian distribution with zero mean.
The interaction has the following properties
\begin{eqnarray}
J_{ij,kl}=-J_{ji,kl}, &  & J_{ij,kl}=-J_{ij,lk}, \nonumber \\
J_{ij,kl}=-J_{kl,ij}^{*}, &  & \overline{|J_{ij,kl}|^{2}}=J^{2}.\label{eq:original_random_coupling}
\end{eqnarray}
%


The original SYK model Eq.~(\ref{eq:original_syk_hamiltonian}) constituted of complex interactions contains 
both amplitude and phase randomness, hence is in principle hard to realize experimentally. 
To overcome this difficulty, we consider a variant model by setting the couplings to be real 
and take values separated by a specific step, i.e.,
\begin{eqnarray}
\tilde{J}_{ij,kl} & = & \textrm{{\bf round}}(J_{ij,kl}/d)\times d,\label{eq:real_discrete_random_coupling}
\end{eqnarray}
where $J_{ij,kl}$ are real Gaussian random couplings distributed as
\begin{eqnarray}
J_{ij,kl}=-J_{ji,kl}, &  & J_{ij,kl}=-J_{ij,lk},\nonumber \\
J_{ij,kl}=-J_{kl,ij}, &  & \overline{|J_{ij,kl}|^{2}}=J^{2},\label{eq:real_random_coupling}
\end{eqnarray}
and the $\text{{\bf round}}$ function returns the value of its argument
rounded to the nearest integer. The Hamiltonian finally becomes
%
\begin{eqnarray}
\mathcal{H}^{\text{dis}} & = & \frac{1}{(2N)^{\frac{3}{2}}}\sum_{i,j,k,l}^{N}\tilde{J}_{ij,kl}
c_{i}^{\dagger}c_{j}^{\dagger}c_{k}c_{l} - \mu \sum_i c_i^\dagger c_i.
\label{eq:real_discrete_syk_hamiltonian}
\end{eqnarray}
From now on, we denote the original SYK Hamiltonian Eq.~(\ref{eq:original_syk_hamiltonian})
as complex-SYK model, and the variation of original SYK model where the
couplings are replaced with real Gaussian random numbers is referred
to as real-SYK model~\cite{real-SYK-modify}. Meanwhile, the Hamiltonian
in Eq.~(\ref{eq:real_discrete_syk_hamiltonian}) is referred as discrete-real-SYK
model, which recovers the continuous-real-SYK model as the discrete distance
approaches to zero. In the opposite limit of large discrete distance $d$, 
most of the random numbers are rounded to zero, making the Hamiltonian 
returns into a non-interacting model. Next, we use exact diagonalization 
to study the three models and compare results for various properties.
As the outcome of the complex-SYK model and the continuous-real-SYK model
are very close for all system sizes we have considered, in the following discussion 
we do not distinguish the two cases and refer to them as continuous (cont.) models.

\section{Ground state and level spectrum}
\label{sec:zeroT}

A remarkable property of the complex-SYK model is the non-Fermi liquid
ground state with volume law of bipartite entanglement, which is the
first quantity we explore here. The problem we are interested in is to what extent 
does the basic property of the complex-SYK model remains. 
Firstly, we focus on the zero-temperature properties,
so only a sector of Hilbert space with fixed particle number should
be taken into consideration. For numerical convenience, we choose the 
half-filling sector and carry out calculation for system size $N$ a multiple of 4.
\begin{figure}[t]
\includegraphics[width=0.48\textwidth]{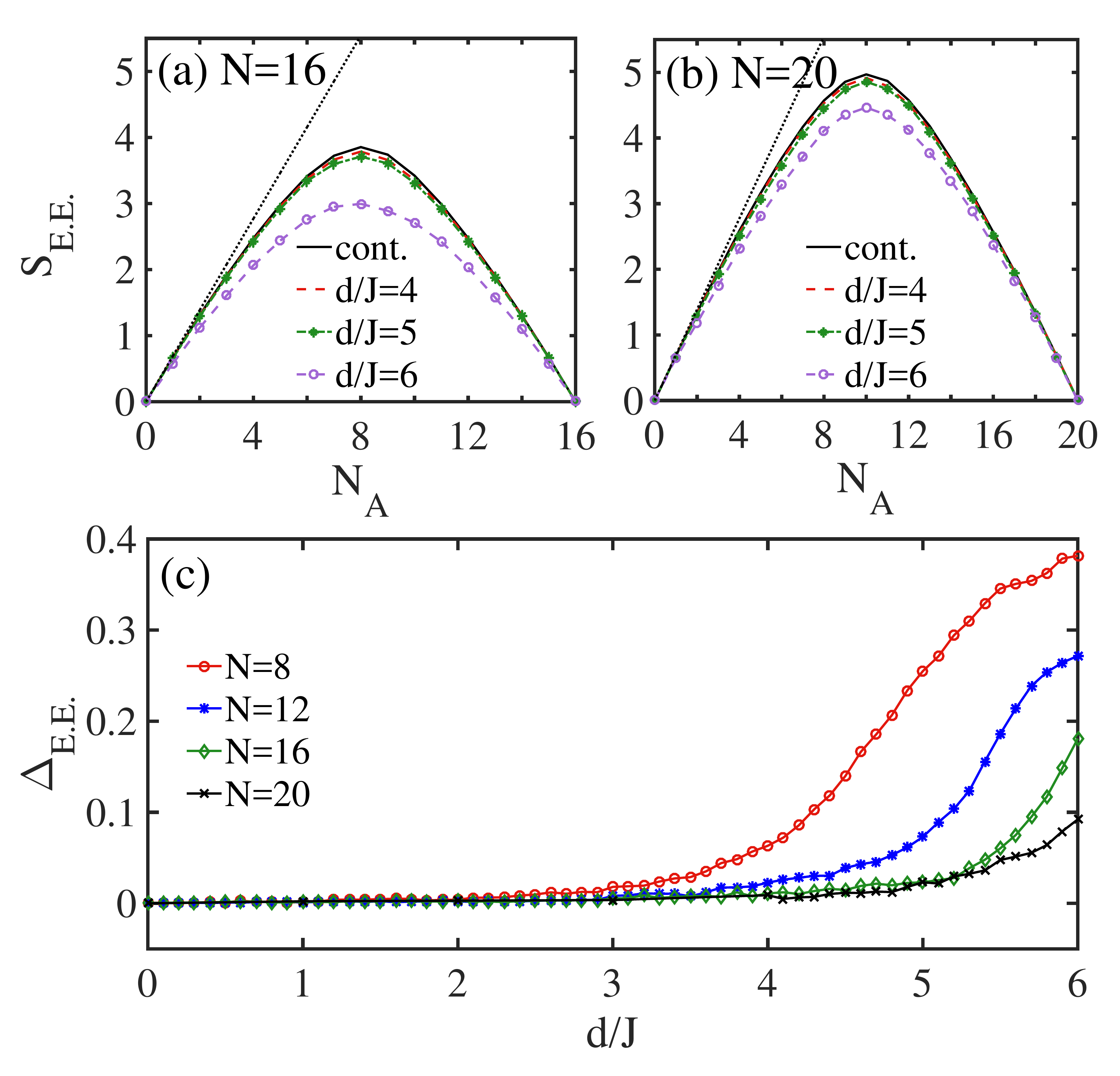} 
\caption{(Color online) Entanglement entropy ($S_{\rm E.E.}$) of a subsystem 
of size $N_A$ in a lattice of size (a) $N=12$ and (b) $N=16$. 
The relative difference $\Delta_{\rm E.E.} $between the discrete-real-SYK model and 
the continuous-SYK model (both real and complex) for different choices of 
discrete distance $d/J$ are shown in (c).}
\label{fig:ground-state-entanglement}
\end{figure}

In Fig.~\ref{fig:ground-state-entanglement}, we present the ground-state 
entanglement entropy for the discrete-real-SYK model by varying the discrete 
separation $d$. The entanglement entropy ($S_{\text{E.E.}}$) is obtained by calculating
the von Neumann entropy of the reduced density matrix of subsystem denoted as $A$,
\begin{eqnarray}
S_{\text{E.E}} & = & -\text{Tr}(\rho_{\text{A}}\ln\rho_{\text{A}}),\label{eq:entanglement-entropy}\\
\rho_{\text{A}} & = & \text{\text{Tr}}_{\bar{\text{A}}}(|G\rangle\langle G|),\label{eq:reduced-dentisy-matrix}
\end{eqnarray}
where $\{A,\bar{A}\}$ partitions the spatial freedoms of the lattice.
Specifically, we choose $N_{\text{A}}$ consecutive sites on the left as the
subsystem, $\rho_{\text{A}}$ is the corresponding reduced density matrix, 
and $|G\rangle$ the ground state of Hamiltonian Eq.~(\ref{eq:real_discrete_syk_hamiltonian}).
As shown by black solid lines in Figs.~\ref{fig:ground-state-entanglement}(a) and \ref{fig:ground-state-entanglement}(b),
the entanglement entropy density of continuous-SYK model (both complex and real)
comes between $\ln2$ (dotted line) and the analytic result 0.464848 in zero-temperature limit~\cite{Fu2016}. 

We then compare the result with the outcome of discrete models. 
When the separation  $d/J$ between the possible strengths of random interactions is small, 
the entanglement entropy obtained from the discrete model gets reduced from
the continuous model, with a very slight deviation for all lattice sizes up to the largest 
value we have considered $N=20$, as shown in Figs.~\ref{fig:ground-state-entanglement}(a) 
and \ref{fig:ground-state-entanglement}(b). With increasing $d/J$, the discrepancy becomes 
more evident and the entanglement entropy is further reduced. This tendency is naturally 
expected since the system approaches to a non-interacting model in the 
limiting case of large $d/J$. However, an interesting and important finding
is that the deviation between the discrete and continuous models is 
not linearly dependent on $d/J$. To further quantify this observation,
we calculate the relative difference between curves of the discrete and 
continuous models by summing over lattice points except the end points where
the entanglement entropy is trivially zero
\begin{eqnarray}
\label{eq:relative difference}
\Delta_{\rm E.E.} \equiv  \frac{1}{N_A} \sum_{N_A} \frac{\left \vert S_{\rm E.E.}^{\rm (disc.)} 
- S_{\rm E.E}^{\rm (cont.)} \right \vert}{S_{\rm E.E}^{\rm (cont.)}}.
\end{eqnarray}
The superscript disc. and cond. stand for results from the discrete 
and continuous models, respectively.
As shown in Fig.~\ref{fig:ground-state-entanglement}(c), in the regime 
where the distance $d/J$ is small, the relative difference nearly remains zero,
indicating that the continuous model can be well approximated by the discrete models.
When the separation exceeds a critical value, the discrepancy is rapldly
enhanced and presents a transition-like behavior. The critical value 
also increases with the system size, suggesting that for larger systems, 
a fewer number of random interaction strengths are required to mimic the 
SYK model. Since the system is reduced to a non-interacting model 
with $d/J \to \infty$, one would expect that the critical threshold should 
converge to an asymptotic value in the thermodynamic limit, which, however,
can not be extracted from our numerical results since the calculation is 
very costly and can only be restricted to systems with $N \le 20$.

\begin{figure}[t]
\includegraphics[width=0.48\textwidth]{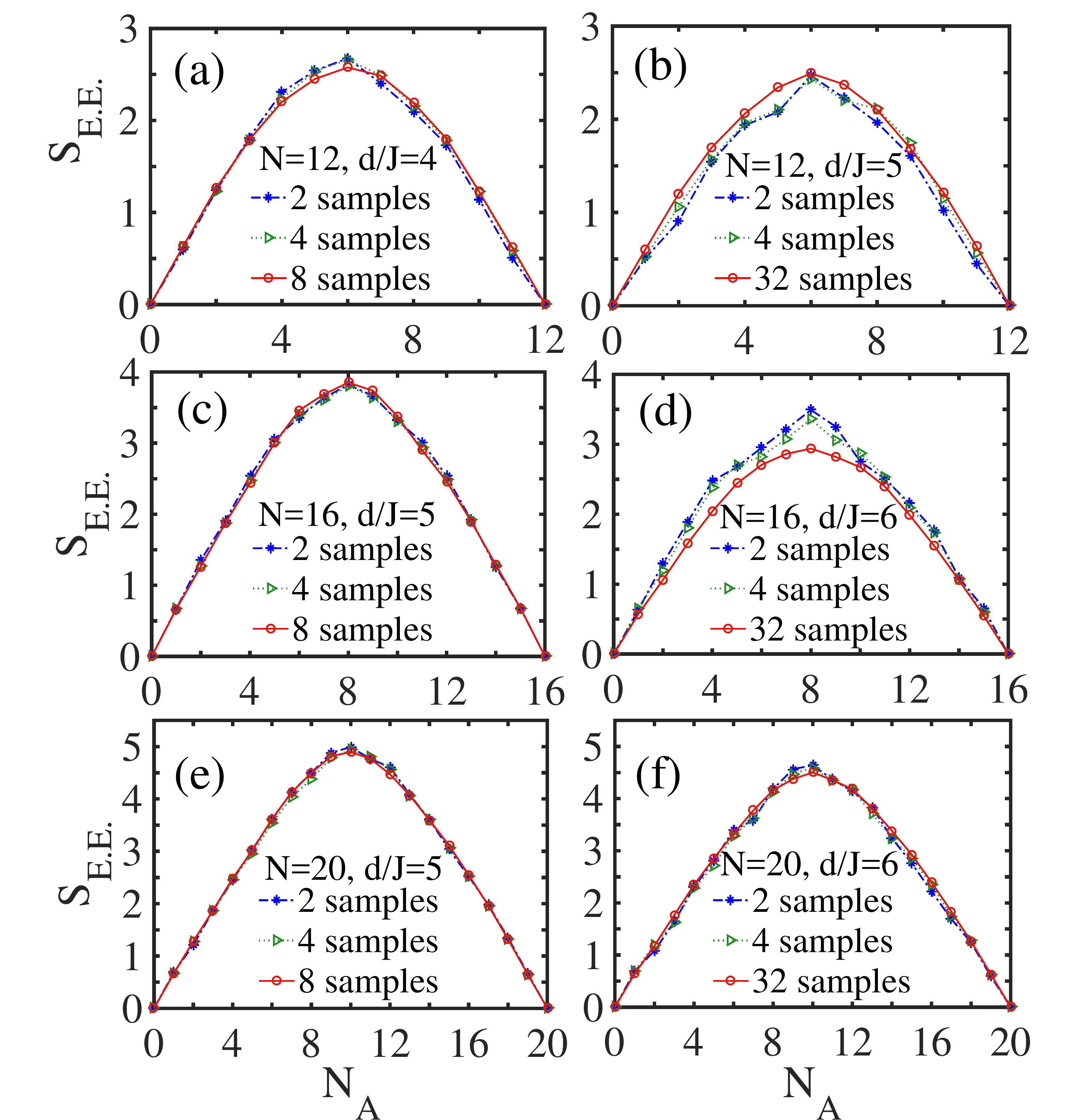} 
\caption{(Color online) Convergence of the entanglement entropy
before (left column) and after (right column) the transition. For $d/J$ 
larger than the critical value, more samples are required to approach the 
averaged value. This observation suggests that the system is no longer fully 
thermalized, as it does in a chaotic phase with small $d/J$.}
\label{fig:entanglement-entropy-convergence}
\end{figure}
The results shown in Fig.~\ref{fig:ground-state-entanglement}
are averaged over a few tens sampling systems.
To check convergence, we exhibit the averaged results for different
sampling numbers, as illustrated in Fig.~\ref{fig:entanglement-entropy-convergence}.
Comparing to models with only short-range interactions,
the ground state of SYK model is believed to be thermalized, which
means that the results for any random samplings are almost the same
in a sufficiently large system. Indeed, for the continuous-real-SYK model
with 12 or more sites, the numerical result of entanglement entropy of one specific 
sampling is already very close to the converged average.
For discrete models, as the interaction can only choose from a finite 
set of values, with a large probability to be zero in particular, 
the variance of a given sampling is more evident. 
When the separation $d/J$ is less than the critical value, 
the convergence can still be quickly obtained for a few samplings, 
as depicted in Figs.~\ref{fig:entanglement-entropy-convergence}(a), 
\ref{fig:entanglement-entropy-convergence}(c), and \ref{fig:entanglement-entropy-convergence}(e). 
However, in the regime above the threshold as in Figs.~\ref{fig:entanglement-entropy-convergence}(b),
\ref{fig:entanglement-entropy-convergence}(d) and \ref{fig:entanglement-entropy-convergence}(f),
the convergence becomes fairly slow and one has to average over more than a few tens of samples. 
This result implies the existence of a threshold as a function of system size
in the perspective of ETH, in consistence with the trend extracted from entanglement entropy 
shown in Fig.~\ref{fig:ground-state-entanglement}(c).

\begin{figure}[h]
\includegraphics[width=0.48\textwidth]{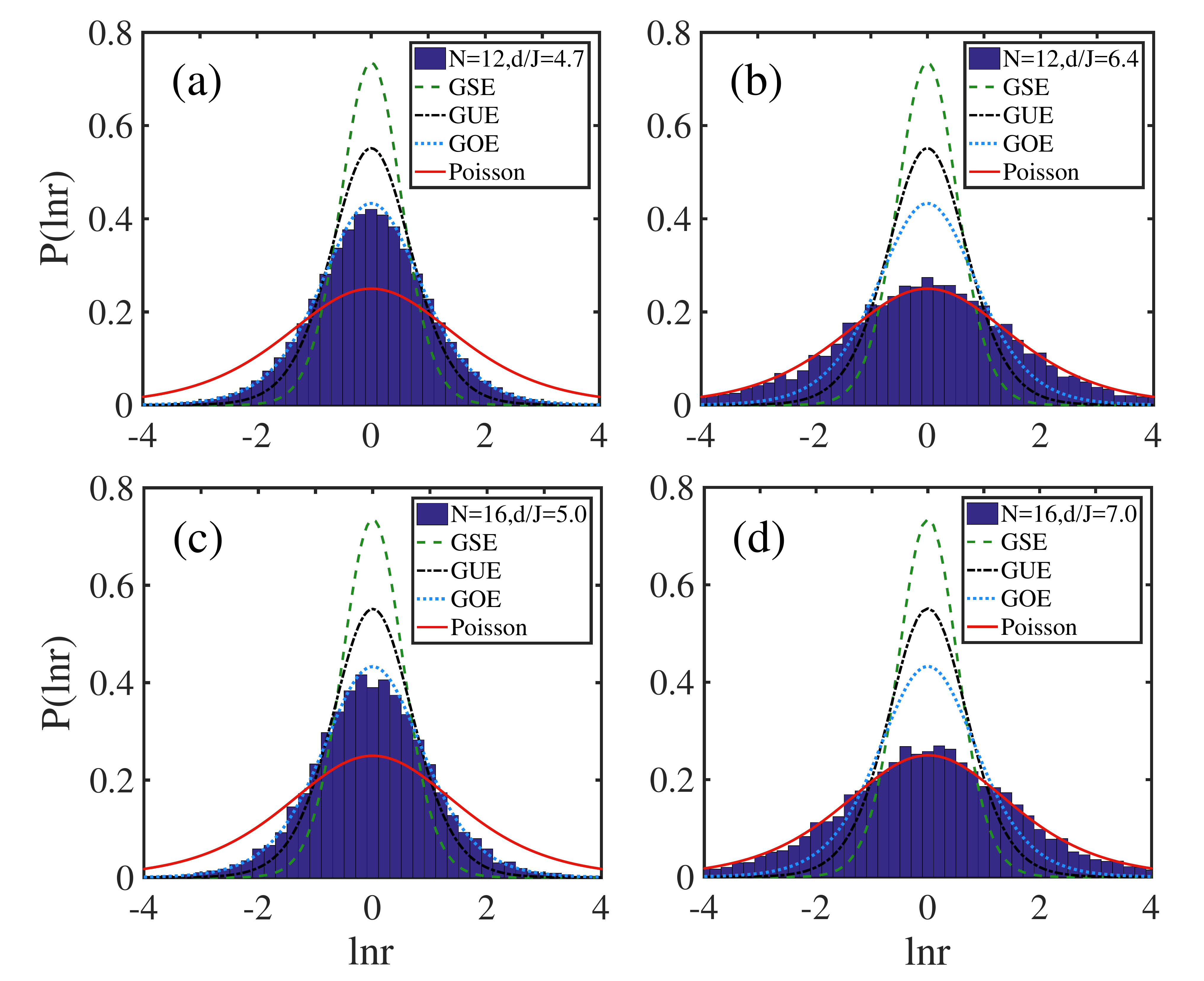}
\caption{(Color online) Level statistics (blue histogram) of the discrete model before (left column) 
and after (right column) the transition. The system clearly evolves from a chaotic state with Wiger--Dyson 
distribution (GOE) at small $d/J$ to an integrable state with Poisson distribution at large $d/J$. }
\label{fig:spectrum}
\end{figure}

To characterize the nature of this phase transition, we then analyze the energy spectrum of the system.
For the original SYK model, the ground state is the maximally chaotic non-Fermi liquid phase, with an 
energy spectrum obeying one of the three chaotic ensembles depending on the number of 
particles.~\cite{Garcia2016,You2017}
The three ensembles correspond to quantum Hamiltonians of 
random matrices whose entries are real (GOE), complex (GUE), or quaternionic (GSE) variables,
For the specific filling factor considered in the present case, the many-body energy spectrum 
of the original SYK model falls into the GOE class. By ordering the energy levels from bottom to top, 
we can define the ratio between successive spacings~\cite{oganesyan_huse_07,Atas-13}
\begin{eqnarray}
\label{eqn:r}
r_n = \frac{E_{n+1} - E_{n}}{E_{n+2} - E_{n+1}},
\end{eqnarray}
which is predicted to be distributed according to the following function
\begin{eqnarray}
\label{eqn:Pr}
P(r) = \frac{1}{Z} \frac{(r+r^2)^\beta}{(1+r+r^2)^{1+3\beta/2}}
\end{eqnarray}
with Dyson index $\beta =1$ and normalization constant $Z= 8/27$.

In Fig.~\ref{fig:spectrum}, we show the histogram of $r_n$ for the discrete-real-SYK model
at typical separations $d/J$ before and after the transition for systems of size $N = 12$ and $16$. 
For the cases of small $d/J$ as in Figs.~\ref{fig:spectrum}(a) and \ref{fig:spectrum}(c), 
the distribution $P(r)$ obeys the prediction for GOE, indicating that the system is in a quantum 
chaotic phase. After crossing the transition point, however, the level distribution approaches to 
a Poisson statistics $P(r) = e^{-r}$, which corresponds to an integrable quantum system. 
A similar chaotic-integrable transition is investigated in another variant of SYK model with 
an additional one-body infinite-range random interaction.~\cite{Garcia2018} Here, we show that 
such a transition can be driven by simply reducing the randomness of the two-body interaction, 
without introducing any other mechanisms. 

\begin{figure}[t]
\includegraphics[width=0.5\textwidth]{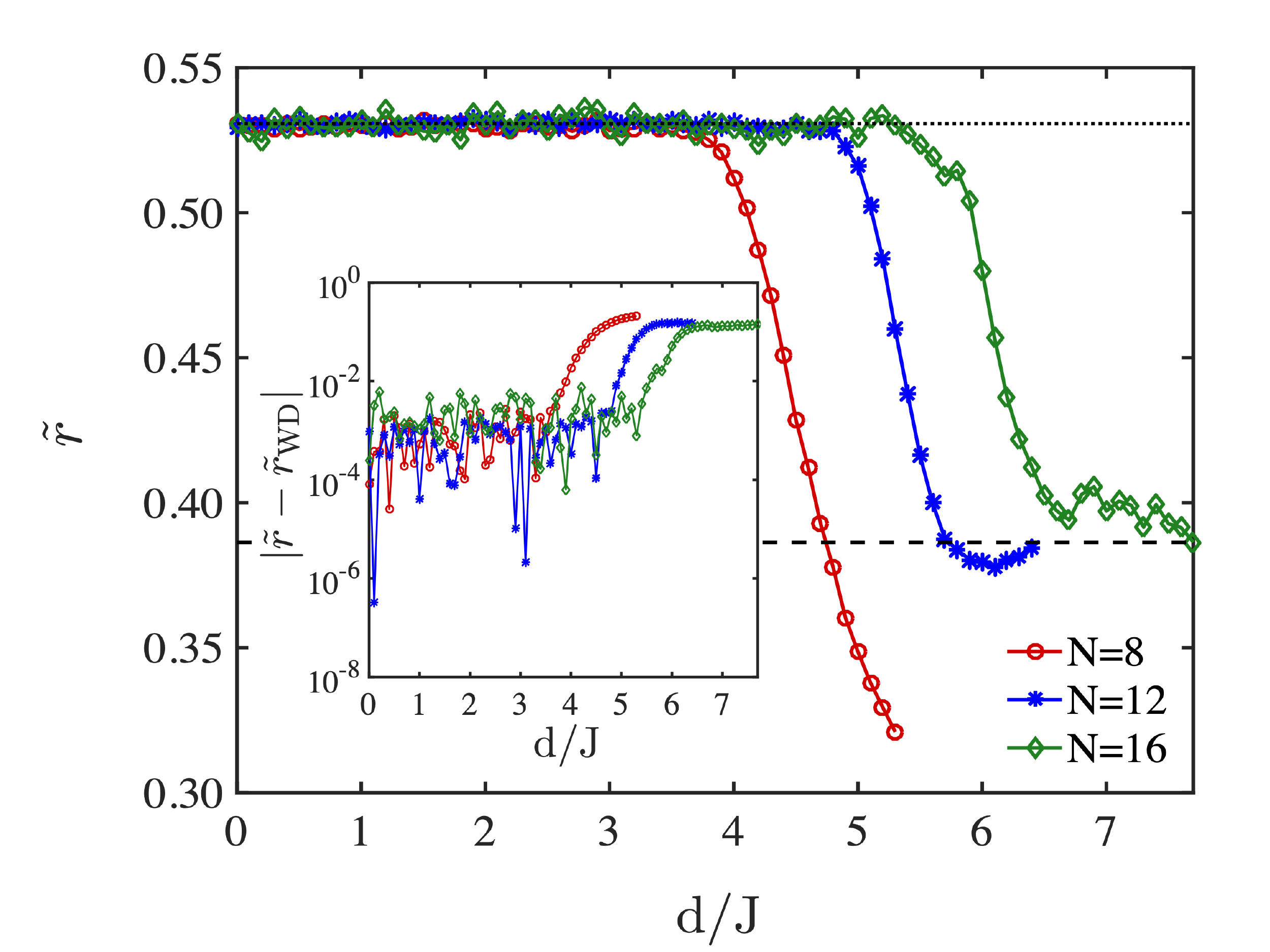} 
\caption{(Color online) Average ratio of adjacent energy gaps as a function of separation
$d/J$. Below the critical threshold, the results for all system sizes recover the value 0.5307 
for a chaotic state. By crossing the transition point, the average ratio starts to drop and saturates
to the limiting value of 0.3863 of Poisson statistics. Inset depicts the same plots in the logarithmic scale,
showing that the deviation from the chaotic state is indeed a sharp transition rather than an exponential 
enhancement.}
\label{fig:distribution}
\end{figure}

Another characteristic quantity of the level statistics is the average ratio between the smallest
and the largest adjacent energy gaps~\cite{oganesyan_huse_07}
\begin{eqnarray}
\label{eqn:r-ratio}
{\tilde r}_n =\frac{\min[\delta^E_n,\delta^E_{n-1}]}{\max[\delta^E_n,\delta^E_{n-1}]}.
\end{eqnarray}
Here, $\{E_n\}$ is the ordered list of energy levels and 
$\delta^E_n = E_n - E_{n-1}$ is the separation between two adjacent eigenstates. 
From Fig.~\ref{fig:distribution}, we find that when $d/J$ is smaller than the critical threshold 
for a give system size, the average ratio recovers the result 
$\langle {\tilde r} \rangle \equiv r_{\rm WD} \approx 0.5307$ (dotted line) of a Wigner--Dyson distribution for GOE. 
If the separation $d/J$ exceeds the critical value, the average ratio ${\tilde r}$ deviates significantly 
from that of a thermal state, and approaches to the limiting value of $2 \ln 2 -1 \approx 0.3863$ (dashed line)
for a Poisson distribution in large enough systems. For the smallest system $N=8$, the result of ${\tilde r}$
goes below the Poisson limit and does not show any clear saturation for the largest value of $d/J$ we have tried. 
In fact, for such a small system with $d/J\gtrsim5$, most of the interaction terms ($\gtrsim 98.7\%$) are tuned off 
and the system presents many nearly degenerate energy levels, which compromise the definition and 
determination of ${\tilde r}$. We also plot the difference of ${\tilde r}$ from the Wigner--Dyson 
result ${\tilde r}_{\rm WD}$ in a logarithmic scale in the inset of Fig.~\ref{fig:distribution}, from which one 
can conclude that the deviation from the chaotic state is indeed a quantum phase transition. 
The transition points for specific lattice sizes agree well with the result of entanglement entropy 
as in Fig.~\ref{fig:ground-state-entanglement}(c). 

%

\section{finite temperature properties}

Next, we probe the tolerance to discretization at finite temperature.
For a physical quantity $f$, we define the relative difference from the continuous result as
\begin{eqnarray}
\Delta_f \equiv
\frac{\int_{a}^{b}dx\left [f^{\rm (disc.)}(x)-f^{\rm (cont.)}(x) \right]}
{\int_{a}^{b}dx f^{\rm (cont.)}(x)},
\label{eqn:relativediff}
\end{eqnarray}
where $a$ (b) is the lower (upper) bound of the parameter range,
and $f^{\rm (disc.)}$ ($f^{\rm (cont.)}$) is a function to be evaluated for the 
discrete (continuous) SYK model.

\begin{figure}[t]
\includegraphics[width=0.48\textwidth]{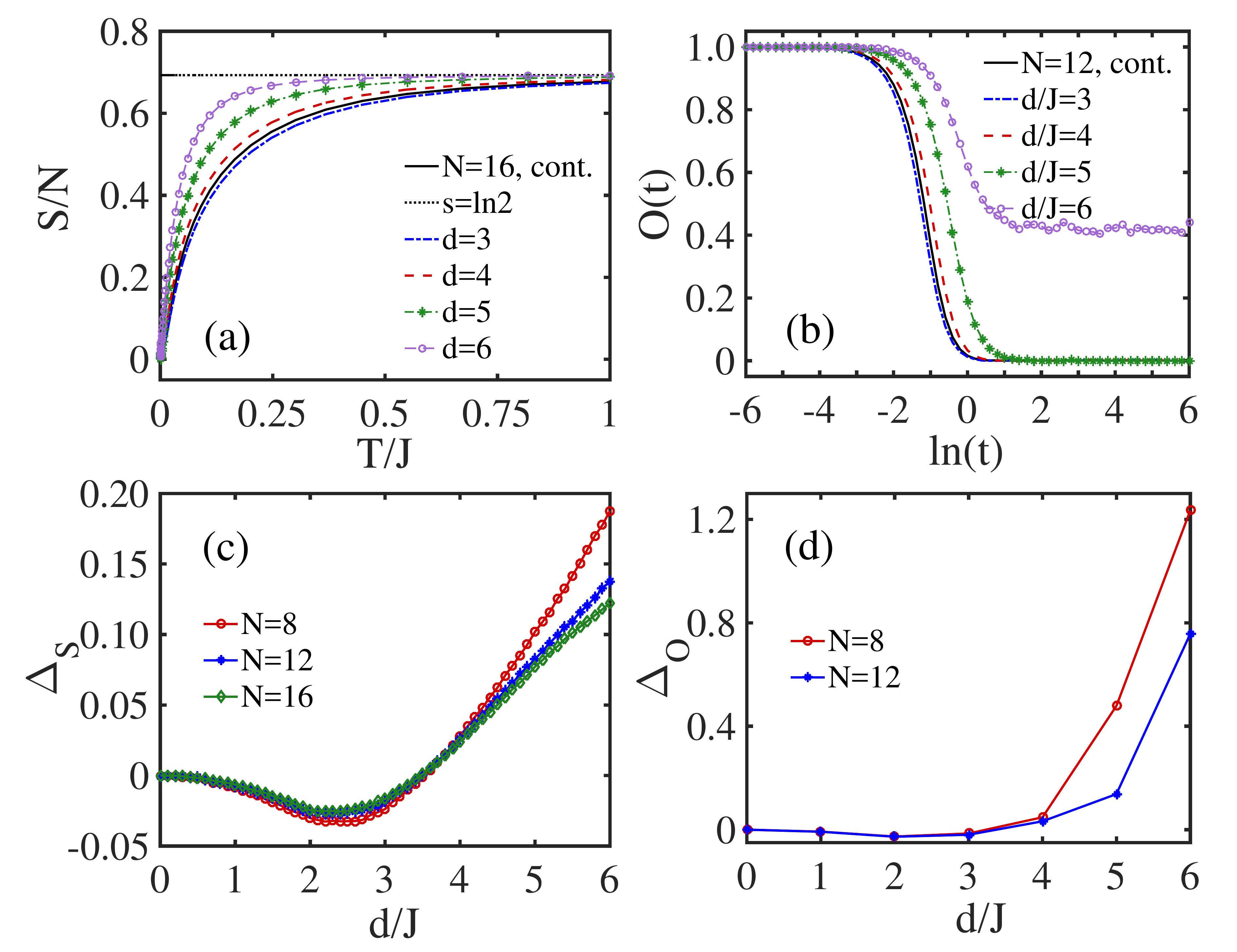} 
\caption{(Color online) (a) Thermal entropy density as a function of temperature
with different discretization of couplings for $N=16$, and (c) the
relative differences from continuous model. Results for all configurations
approach the high temperature limit $S/N=\ln2$, however, analytical
result for the zero-temperature limit can not be obtained for finite size calculation.
(b) OTOC as a function of time on logarithmic scales at
$T/J$ =1 with different discretization of couplings for $N=12$, and (d)
its relative difference from the continuous model.}
\label{fig:entropy-OTOC}
\end{figure}

We first examine the thermodynamic entropy, which can be expressed as 
\begin{eqnarray}
\frac{S}{N}=\frac{\langle E\rangle/T+\ln\mathcal{Z}}{N}.
\label{eq:thermal-entropy-density}
\end{eqnarray}
Here, the partition function $\mathcal{Z}$ reads $\mathcal{Z}=\sum_{n}e^{-E_{n}/T}$
with $E_{n}$ the eigenvalues for the canonical ensemble, 
and $\langle E\rangle= (1/\mathcal{Z})\sum_{n}E_{n}e^{-E_{n}/T}$ is the average energy. 
Again, we find that the discrete-real-SYK model can mimic the original model 
provided that the separation $d/J$ is smaller than the critical value. 
However, the transition is not as sharp as in Figs.~\ref{fig:ground-state-entanglement}
and \ref{fig:distribution}, and does not show a clear enhancement with 
system size. The relative difference becomes significant when $d/J \gtrsim 4$ for 
both $N = 12$ and 16, as shown in Figs.~\ref{fig:entropy-OTOC}(a) and \ref{fig:entropy-OTOC}(c). 
Another observation is that the relative difference first becomes negative
and then increases with $d$ on the rise. Therefore, in the perspective
of relative difference, there is a specific value for $d/J \approx 3.5$
which can mostly recover the original SYK model.

Finally, we analyze the exponential decay of the out-of-time-ordered correlation (OTOC) 
which is closely related to the quantum butterfly effect.
The decay rate $\lambda_{L}$ is known as Lyapunov exponent in classical
chaos theory and depicts the strength of chaos. The upper bound of $\lambda_{L}$
is determined in the strong-coupling limit as $2\pi/\beta$ with $\beta$ 
the inverse temperature, and is argued to be reached in the SYK model for 
strongly interacting Majorana fermions in the thermodynamic limit, 
as well as in the theory of Einstein gravity on the $\text{AdS}_{2}$ horizons 
of blackholes.~\cite{Hosur2016, Fu2016}
However, in finite-size calculations, the theoretical results can not be reproduced 
in the strong coupling limit and $\lambda_{L}$ is not sensitive to $\beta J$ at finite
temperatures~\cite{Fu2016}. Here, we define a normalized OTOC as
\begin{eqnarray}
{\cal O}(t) & = & -\frac{\text{Re}\langle A(t)B(0)A(t)B(0)\rangle_{\beta}}{\langle A(0)A(0)\rangle_{\beta}\langle B(0)B(0)\rangle_{\beta}},
\end{eqnarray}
where $A$ and $B$ are Majorana operators
\begin{eqnarray}
A & = & c_{1}+c_{1}^{\dagger},\\
B & = & c_{2}+c_{2}^{\dagger}.
\end{eqnarray}
Since the Lyapunov exponent is not sensitive to the coupling strength, we choose 
a moderate value $\beta J=1$ and investigate the influence of discretization as shown 
in Fig.~\ref{fig:entropy-OTOC}.

In line with expectations, the scrambling of OTOC almost keeps unchanged with
small $d$. However, when $d/J$ goes beyond a similar threshold, the deviation becomes prominent
as shown in Fig.~\ref{fig:entropy-OTOC}(b). Importantly, we find that ${\cal O}(t)$ evolves to a finite value 
rather than zero in the long-time limit. This observation can be understood by noticing that as the
couplings become more discretized, most of them are rounded to zero, 
in which case the Hamiltonian commutes with the Majorana operators.
If the commutative case takes a non-zero measure among the samplings,
information can be preserved during forward and backward evolution and 
${\cal O}(t)$ does not converge to zero. In Fig.~\ref{fig:entropy-OTOC}(d), we also show 
the relative difference between the discrete and continuous models as defined
in Eq.~(\ref{eqn:relativediff}). A transition-like behavior is also observed when $d/J$ 
exceeds a critical value, which seems to increase with system size.

\section{Summary and Discussion}
In this paper, we numerically study a simplified variant of the Sachdev-Ye-Kitaev (SYK) model 
where the all-to-all random two-body interaction can only take real values
which are discretely separated by a finite distance $d$. Using exact diagonalization,
we numerically solve this discrete model and compare results with the original SYK model
for systems of size up to $N=20$ at half filling. By increasing the separation $d$, or equivalently
reduces the randomness of the interaction, we observe a quantum phase transition from
chaotic to integrable states, which can be signatures in various physical quantities including 
entanglement entropy, level spectrum, thermodynamic entropy, and out-of-time-ordered correlation. 
The transition point increases with system size, indicating that for larger systems, a chaotic state
can be stabilized with a relatively weak randomness. These results extend our knowledge about 
the SYK model, and about the more general topic of quantum chaotic system.

More importantly, our study paves the way towards the experimental simulation of the SYK model
by significantly relaxing the requirement of the interaction. First, we show that the interaction 
does not to be of infinite range. In fact, most of the inter-particle combinations can be assumed 
to be non-interacting while the system still reserves the behavior of the SYK model. For instance, 
in a system of size $N=8$, a discrete model with separation of $d/J = 3$ can mimic the original SYK
model in all aspects, where only $~9.5\%$ of the total $8! = 1168$ possible interaction terms are non-zero.
As the critical transition point increases with the lattice size, this condition is further relaxed for 
larger systems. Another advantage of our model is that the interaction strength can only take 
a few possible choices, rather than a completely random distribution. In previous experimental proposals,
much effort has been devoted to achieve the desired randomness by either fabricating an irregular
sample~\cite{Pikulin017, Chen2018} or applying multiple laser beams~\cite{Danshita2017}, which are 
all experimentally challenging. In a discrete model with $d/J = 3$, one only needs to take into 
account at most four possible choices of $\pm d/J$ and $2\pm d/J$, with the probabilities of 
$9.0\%$ and $0.45\%$, respectively. This can significantly reduce the complexity of experiments. 

\begin{acknowledgments}
We thank M. Tezuka and N.-H. Tong for helpful discussion. 
This work is supported by the National Natural Science Foundation
of China (Grants No. 11434011, 11522436, 11774425, 11704029), 
the National Key R$\&$D Program of China (Grants No. 2018YFA0306501),
the Beijing Natural Science Foundation (Grant No. Z180013), 
and the Research Funds of Renmin University of China 
(Grants No. 16XNLQ03 and 18XNLQ15).
\end{acknowledgments}


\begin{thebibliography}{10}
\bibitem{Sherrington1975} D. Sherrington and S. Kirkpatrick, 
Phys. Rev. Lett. \textbf{35}, 1792 (1975).

\bibitem{Sachdev1993} S. Sachdev and J. Ye, 
Phys. Rev. Lett. \textbf{70}, 3339 (1993).

\bibitem{Parcollet1999} O. Parcollet and A. Georges, 
Phys. Rev. B \textbf{59}, 5341 (1999).

\bibitem{Georges2000} A. Georges, O. Parcollet, and S. Sachdev, 
Phys. Rev. Lett. ${\bf 85}$, 840 (2000).

\bibitem{Georges2001} A. Georges, O. Parcollet, and S. Sachdev, 
Phys. Rev. B ${\bf 63}$, 134406 (2001).

\bibitem{Arrachea2002} L. Arrachea and M. J. Rozenberg, 
Phys. Rev. B \textbf{65}, 224430 (2002).

\bibitem{Camjayi2003} A. Camjayi and M. J. Rozenberg, 
Phys. Rev. Lett. \textbf{90}, 217202 (2003).

\bibitem{Kitaev2015} A. Y. Kitaev, A simple model of quantum holography,
in KITP strings seminar and Entanglement 2015 program, UC Santa Barbara,
Santa Barbara USA, 12 February, 7 April and 27 May 2015, \url{http://online.kitp.ucsb.edu/online/ entangled15/}.

\bibitem{Sachdev2015} S. Sachdev, 
Phys. Rev. X 5, 041025 (2015).

\bibitem{Maldacena2016} J. Maldacena, S. H. Shenker, and D. Stanford,
J. High Energy Phys. {\bf 8}, 106 (2016).

\bibitem{Hosur2016} P. Hosur, X.-L. Qi, D. A. Roberts, and B. Yoshida,
J. High Energy Phys. {\bf 2}, 4 (2016).

\bibitem{Polchinski2016} J. Polchinski and V. Rosenhaus, 
J. High Energy Phys. {\bf 4}, 1 (2016).

\bibitem{Garcia2016} A. M. Garc{\'i}a-Garc{\'i}a and J. J. M. Verbaarschot,
Phys. Rev. D \textbf{94}, 126010 (2016).

\bibitem{You2017} Y.-Z. You, A.W. W. Ludwig, and C. Xu, 
Phys. Rev. B \textbf{95}, 115150 (2017).

\bibitem{Garcia2018} A. M. Garc{\'i}a-garc{\'i}a, B. Loureiro, A. Romero-berm{\'u}dez, and M. Tezuka, 
Phys. Rev. Lett. \textbf{120}, 241603 (2018).

\bibitem{Scaffidi2019} T. Scaffidi and E. Altman, 
Phys. Rev. B \textbf{100}, 155128 (2019).

\bibitem{Garcia2019-chaos} A. M. Garc{\'i}a-garc{\'i}a, T. Nosaka, D. Rosa, and J. J. M. Verbaarschot, 
Phys. Rev. D \textbf{100}, 26002 (2019).

\bibitem{Zhang2018} P. Zhang and H. Zhai, 
Phys. Rev. B \textbf{97}, 201112 (2018).

\bibitem{Banerjee2017} S. Banerjee and E. Altman, 
Phys. Rev. B \textbf{95}, 134302 (2017).

\bibitem{Bi2017} Z. Bi, C.-M. Jian, Y.-Z. You, K. A. Pawlak, and C. Xu, 
Phys. Rev. B \textbf{95}, 205105 (2017).

\bibitem{Can2019-phase-transition} O. Can and M. Franz, 
Phys. Rev. B \textbf{100}, 45124 (2019).

\bibitem{Altland2019} A. Altland, D. Bagrets, and A. Kamenev, 
Phys. Rev. Lett. \textbf{123}, 106601 (2019).

\bibitem{Guo2019} H. Guo, Y. Gu, and S. Sachdev, 
Phys. Rev. B \textbf{100}, 045140 (2019).

\bibitem{Garcia2017} A. M. Garc{\'i}a-Garc{\'i}a and J. J. M. Verbaarschot,
Phys. Rev. D \textbf{96}, 066012 (2017).

\bibitem{Jian S.-K2017} S.-K. Jian and H. Yao, 
Phys. Rev. Lett. \textbf{119}, 206602 (2017).

\bibitem{Jian C-.M2017} C.-M. Jian, Z. Bi, and C. Xu, 
Phys. Rev. B \textbf{96}, 115122 (2017).

\bibitem{Song2017}X.-Y. Song, C.-M. Jian, and L. Balents, 
Phys. Rev. Lett. \textbf{119}, 216601 (2017).

\bibitem{Chen2017} X. Chen, R. Fan, Y. Chen, H. Zhai, and P. Zhang,
Phys. Rev. Lett. \textbf{119}, 207603 (2017).

\bibitem{Trajewski2019} T. Krajewski, M. Laudonio, R. Pascalie, and A. Tanasa, 
Phys. Rev. D \textbf{99}, 126014 (2019).

\bibitem{Iyoda2018} E. Iyoda, H. Katsura, and T. Sagawa, 
Phys. Rev. D \textbf{98}, 086020 (2018).

\bibitem{Garcia2019-mbl} A. M. Garc{\'i}a-Garc{\'i}a and M. Tezuka, 
Phys. Rev. B \textbf{99}, 054202 (2019).

\bibitem{Pikulin017} D. I. Pikulin and M. Franz, 
Phys. Rev. X \textbf{7}, 031006 (2017).

\bibitem{Chew2017} A. Chew, A. Essin, and J. Alicea, 
Phys. Rev. B \textbf{96}, 121119 (2017).

\bibitem{Chen2018} A. Chen, R. Ilan, F. de Juan, D. I. Pikulin, and M. Franz, 
Phys. Rev. Lett. \textbf{121}, 036403 (2018).

\bibitem{Can2019-transport} O. Can, E. M. Nica, and M. Franz, 
Phys. Rev. B \textbf{99}, 045419 (2019).

\bibitem{Danshita2017} I. Danshita, M. Hanada, andM. Tezuka, 
Prog. Theor. Exp. Phys. {\bf 2017}, 083I01 (2017).

\bibitem{Fu2016} W. Fu and S. Sachdev, 
Phys. Rev. B \textbf{94}, 035135 (2016).

\bibitem{real-SYK-modify} The real-SYK model we denoted here is slightly
different with that in I. Danshita, M. Hanada, and M. Tezuka's work
\cite{Danshita2017}, where the variances of diagonal couplings are
twice as the off-diagonal ones. In the numerical calculation, we do
not find any significant difference between two setups.

\bibitem{oganesyan_huse_07} V. Oganesyan and D. A. Huse, 
Phys. Rev. B {\bf 75}, 155111 (2007).

\bibitem{Atas-13} Y. Y. Atas, E. Bogomolny, O. Giraud, and G. Roux, 
Phys. Rev. Lett. {\bf 110}, 084101 (2013).


\end{thebibliography}
\end{document}